\documentclass[12pt]{article}
\usepackage{booktabs}
\usepackage[margin=2cm,a4paper]{geometry} 
\usepackage{amssymb, amsthm, amsmath}
\usepackage{tikz}
\usepackage{enumerate}

\begin{document}

\newcommand{\fra}[2]{\textstyle{\frac{#1}{#2}}}
\newcommand{\sgn}[1]{\text{sgn}(#1)}
\newcommand{\I}{i}

\theoremstyle{plain}
\newtheorem{thm}{Theorem}[section]
\newtheorem{prop}{Proposition}[section]
\newtheorem{cor}[thm]{Corollary}
\newtheorem{res}{Result}
\newtheorem{rem}{Remark}
\newtheorem{lem}[thm]{Lemma}

\newcommand{\lrr}[1]{{\langle #1 \rangle}_{\mathbb{R}}}
\newcommand{\id}{{\bf 1}}

\newcommand{\lie}[2]{\left[#1,#2\right]}
\newcommand{\beqn}{\begin{eqnarray}\begin{aligned}}
\newcommand{\eqn}{\end{aligned}\end{eqnarray}}

\newcommand{\unsignedpermutation}{
\pgfmatharray{\sp}{0}\let\n\pgfmathresult
\pgfmathparse{360.0/\n}\let\segment\pgfmathresult
\pgfmathparse{\segment/2}\let\shift\pgfmathresult
\def\radius{1.5cm}
\def\labelrad{1.8cm}
\def\regionboundaryin{1.4cm}
\def\regionboundaryout{1.6cm}
\begin{tikzpicture}
\foreach \x in {1,2,...,\n}
{
  \draw[thick] (360-\x*\segment+90:\regionboundaryin)
             --(360-\x*\segment+90:\regionboundaryout);
  \pgfmatharray{\sp}{\x}\let\tmp\pgfmathresult 
  \pgfmathparse{int(\tmp)}\let\tmp\pgfmathresult 
  \node at (360-\x*\segment+90+\shift:\labelrad) {\tmp};
  \pgfmathparse{360-(\x-1)*\segment+90}\let\alpha\pgfmathresult;
  \pgfmathparse{360-(\x-1)*\segment+90-\segment}\let\beta\pgfmathresult;
  \pgfmathgreater{\tmp}{0}\let\decision\pgfmathresult
  \ifnum \decision=1
    \draw[color=black,very thick] (\alpha:\radius) arc (\alpha:\beta:\radius);
  \else
    \draw[color=gray,very thick] (\beta:\radius) arc (\beta:\alpha:\radius);
  \fi
};
\end{tikzpicture}
}
\newcommand{\signedpermutation}{
\pgfmatharray{\sp}{0}\let\n\pgfmathresult
\pgfmathparse{360.0/\n}\let\segment\pgfmathresult
\pgfmathparse{\segment/2}\let\shift\pgfmathresult
\def\radius{1.5cm}
\def\labelrad{1.8cm}
\def\regionboundaryin{1.4cm}
\def\regionboundaryout{1.6cm}
{\begin{tikzpicture}
\foreach \x in {1,2,...,\n}
{
  \draw[thick] (360-\x*\segment+90:\regionboundaryin)
             --(360-\x*\segment+90:\regionboundaryout);
  \pgfmatharray{\sp}{\x}\let\tmp\pgfmathresult 
  \pgfmathparse{int(\tmp)}\let\tmp\pgfmathresult 
  \node at (360-\x*\segment+90+\shift:\labelrad) {\tmp};
  \pgfmathparse{360-(\x-1)*\segment+90}\let\alpha\pgfmathresult;
  \pgfmathparse{360-(\x-1)*\segment+90-\segment}\let\beta\pgfmathresult;
  \pgfmathgreater{\tmp}{0}\let\decision\pgfmathresult
  \ifnum \decision=1
    \draw[color=black,very thick,>=latex,->] (\alpha:\radius) arc (\alpha:\beta:\radius);
  \else
    \draw[color=black,very thick,>=latex,->] (\beta:\radius) arc (\beta:\alpha:\radius);
  \fi
};
\end{tikzpicture}}}

\title{A representation-theoretic approach to the calculation of evolutionary distance in bacteria}

\author{Jeremy G Sumner, Peter D Jarvis, and Andrew R Francis}
\maketitle

\begin{abstract}\noindent
In the context of bacteria and models of their evolution under genome rearrangement, we explore a novel application of group representation theory to the inference of evolutionary history.
Our contribution is to show, in a very general maximum likelihood setting, how to use elementary matrix algebra to sidestep intractable combinatorial computations and convert the problem into one of eigenvalue estimation amenable to standard numerical approximation techniques.
\end{abstract}

\section{Phylogenetics and bacterial evolution}

\medskip
Phylogenetics is the suite of mathematical and computational methods which provide biologists with the means to infer past evolutionary relationships between present-day observed species. 
The usual computational output is an inferred evolutionary tree, together with branch lengths and dates of divergences. 
Input to these methods most typically consist of molecular data such as sequence alignments (for eukaryote DNA, the basic datum is the pattern of nucleotides at each site in the alignment). 
The state of the art is to model molecular evolution as a continuous-time Markov chain on a tree, with the content of each site in the alignment presumed to have evolved under the statistical assumption of being independent and identically distributed (IID). 
Proposed evolutionary histories are inferred using standard statistical frameworks, such as maximum likelihood \cite{felsenstein1981,guindon2010new}, or a Bayesian approach \cite{drummond2007,yang1997bayesian} (in the latter the output is not just a single tree proposal, but rather a posterior distribution on all model parameters, including the evolutionary tree itself). 
These methods are only feasible because of significant modern developments in both stochastic modelling and the almost weekly rise of available computer power.

Historically, more elementary approaches to phylogenetic estimation have included maximum parsimony \cite{farris1970methods,fitch1971toward} and distance-based methods \cite{saitou1987,sneath1973numerical}. 
Maximum parsimony proceeds by asking, given a sequence alignment and a candidate evolutionary tree, what is the least number of changes required to fit the observed nucelotide patterns to the leaves of the tree. 
This method is clearly blind to multiple changes and has long been known to be statistically biased (most notably via the well-studied ``long-branch attraction'' effect \cite{felsenstein1978}).
For this reason, parsimony has somewhat fallen out of favour and has been replaced by model-based approaches. 
On the other hand, distance-based methods are a sensible intermediary, as they can be implemented in a statistically consistent manner (with the assumption of a Markov model on a tree), and computationally are very efficient in comparison to maximum likelihood or Bayesian approaches. 
Indeed, the main difficulty for any phylogenetic method is  the huge number of possible (binary) trees: for $L$ taxa there are $\sim 2^LL!$ trees\footnote{There are $(2L-3)!!=(2L-3)(2L-5)\ldots 3\cdot 1$ rooted binary trees with $L$ leaves.}, and the attractive feature of distance-based methods is that they proceed by clustering and hence can create a sensible candidate tree in $\mathcal{O}(L^3)$  time \cite{felsenstein2004}. 
These include the ever-popular  ``Neighbour-Joining'' algorithm \cite{saitou1987}, and its generalizations ``Neighbour-Net'' \cite{bryant2004} and ``Splits-Tree'' \cite{huson2005}.
The latter methods return a network structure which is close to a tree but allows for the visual representation of signals caused by evolutionary events which deviate from strict vertical descent (for example, ``lateral gene transfer'' events\footnote{Where genes from one species are inserted into the genome of another.}) and/or random noise in the sequence data. 
Importantly, current standard practice for both likelihood and Bayesian approaches is to first use a distance-based method to construct a reasonable starting tree and then use hill-climbing on the likelihood function to attempt to find a better solution using small perturbations to the tree under standard set of moves \cite{bryant2005b}.


At their core, distance-based methods rely on finding a statistically consistent method for calculating the distance between a pair of observed sequences (sometimes referred to as a ``genetic'' distance). 
Solving this problem credibly consists of three parts:
\begin{enumerate}[(i)]
\itemsep-.15em
\item Specify a stochastic model of how molecular sequences evolve randomly, for example a fixed Markov model on a tree; 
\item Define mathematically what is meant by distance relative to the model.
For example, in a continuous-time formulation the distance is most sensibly made to be time itself (although this is not the only choice and may depend on extra additional unknowns such as relative rates of molecular substitutions); 
\item Analyse the model to produce a robust and bias free statistical estimator of the distance from observed data. 
\end{enumerate}
\vspace{-.5em}

A simple example illustrating this process is the well-known ``Jukes-Cantor correction'' \cite{jukes1969} which (i) assumes a Markov model where every nucleotide substitution occurs with the equal rate $r$; (ii) defines distance to be the time passed; and (iii) corrects for unobserved multiple changes in the raw number of observed changes $\Delta$ between two sequences of length $N$ (also known as the Hamming distance).
This is achieved with a statistically consistent estimate of time elapsed via the inversion formula $\widehat{T}=-(3/4r)\log(1-\frac{4}{3}\Delta/N)$, which can be shown is the analytic expression for the maximum likelihood estimate of time under the Jukes-Cantor model \cite{gascuel2005}.
Analogous to maximum parsimony, failing to use this correction, for example applying the Hamming distance $\Delta$ directly, can lead to incorrect evolutionary tree inference.


The Jukes-Cantor correction can be generalized to distances consistent with more complicated models such as the Kimura 2ST model \cite{kimura1980}, or even the general Markov model \cite{barry1987} using the so-called ``log-det'' \cite{lockhart1994}. However, as always in any statistical estimation problem, there is a compromise between assuming a complicated model to increase model realism (and hence decreasing bias), with the inevitable increase in variance of the additional parameters requiring estimation in more complex models \cite{burnham2002}.  
This so-called ``bias-variance'' tradeoff is an extremely important consideration in all approaches of phylogenetic estimation methods.

Phylogenetic modelling in the context of bacterial genomes is radically different. 
At the DNA level, evolution in bacteria is far more dynamic than that in eukaryotes, including evolutionary events that violate the assumption of vertical descent.
These include genome rearrangements (relocation of genes within the genome), deletions of large segments of the genome, and lateral gene transfer between species. 
This fluidity makes the methods described above -- the observation of substitutions at single nucleotide sites to infer evolutionary distances -- highly problematic for bacteria. 
A major practical obstruction is that accurate sequence alignment of bacterial genomes at individual nucleotide sites is unattainable. 
For instance, it is often the case that genomes within the same bacterial species share as little as 40\% of the same genes (the situation with \emph{E.coli} is discussed in \cite{welch2002extensive}).


The occurrence of large-scale rearrangements deems the previously described classic phylogenetic approaches (where individual nucleotide sites are modelled under an IID assumption) rather inappropriate for bacteria, and a more coarse-grained approach is required. 
In this context, it is sensible to divide bacterial genomes into ``regions'' (these can be genes or clusters of genes; essentially any component which is sufficiently conserved such that identification of the region is plausible), and then compare similarity between genomes by comparing firstly how many regions the bacteria have in common, and secondly the relative arrangement of the regions on the genome. 

A major drawback for existing models of bacterial evolution is the absence of a statistically consistent distance estimator analogous to the Jukes-Cantor correction.
In principle there is no difficulty for, as we will describe in the next section, it is straightforward to develop a stochastic model of genome rearrangement and produce an estimator for evolutionary distance which is statistically consistent with the model.
However, as we will also discuss, in practice the calculations involved are of factorial complexity (in the number of regions) and, to date only provisional approaches have been explored.
These approaches either compromise on model realism (making strong assumptions regarding the types of rearrangements that can occur), or compromise on statistical consistency, or both. 
For instance, the means to compute, in  polynomial time, the minimal number  of ``signed inversions'' required to convert one genome into another was provided in \cite{hannenhalli1999transforming} (this was improved to linear time in \cite{bader2001linear}).
Comparing to the DNA case, this is analogous to computing a Hamming (or edit) distance, and is usually referred to as ``minimal distance'' in the context of genome rearrangement models.
The results given in \cite{hannenhalli1999transforming} are however limited by strong assumptions on the genome rearrangements that may occur (specifically, each signed inversion is assumed to be equally likely  --- an assumption that is known to be violated, see \cite{darling2008dynamics} for example).
Earlier, \cite{sankoff1997median} presented the so-called \emph{breakpoint} distance.
This provides a minimal distance between two genomes that is very easy to calculate but has the disadvantage of not being closely aligned to a specific model of genome rearrangement.
Using simulation, an empirically-derived correction to the breakpoint distance was presented in \cite{wang2006}, and in \cite{serdoz2016maximum} we explored the importance of specifying a stochastic model of rearrangements, explicitly showing there are cases where the maximum likelihood estimate of time elapsed and the minimal distance produce contradictory answers.


In this paper we work with a model based on a simple stochastic process that allows for an arbitrary set of genome rearrangements.
Our original contribution is to explore the efficient calculation of estimates of elapsed evolutionary time under this model using methods from group representation theory.
We discuss how to convert the problem into the calculation of group characters and hence show the core computational difficulty reduces to an eigenvalue problem.
This approach has an immediate pay off in the reduction of computational complexity from factorial complexity to the square root of factorial complexity (as a conservative estimate).
Although this is a useful theoretical observation it of course does not solve the computational issues inherent in the problem.
Our broader goal is to use these ideas to produce the means to find \emph{numerical approximations} to maximum likelihood estimates of elapsed time, without the need to make unrealistic modelling assumptions.
Ultimately we hope our approach will lead to computationally efficient, usable software for inferring bacterial phylogenetic trees under rearrangement models.

Our mathematical methods are nothing more than undergraduate level linear algebra and standard results from the representation theory of finite groups (we consider \cite{sagan2001} an excellent reference for the later).
It is worth noting that the results on rearrangement distances presented in \cite{eriksen2004} feature application of the irreducible characters of the symmetric group, and hence we are not the first authors to apply character theory to the problem of genome rearrangement models.
However, the results provided in that work are again highly specialized as it is assumed that each transposition of genome regions is equally likely.
To the best of our knowledge, the work presented in this paper is the first to apply more general methods from representation theory, and certainly the first to do so in the context of likelihood calculations under a general model of genome rearrangement.

In Section~\ref{sec:rearrangementmodels} we discuss the details of the basic mathematical framework for setting up a genome rearrangement model.
In particular, but without loss of generality, we discuss how to deal with the case of ``unsigned'' circular genomes.
In Section~\ref{sec:regrep} we use the representation theory of the symmetric group to find a novel means of counting passages between genomes under a given rearrangement model.
We explore this idea in detail and show how --- through character theory --- the relevant computations reduce to computing the eigenvalues of an operator on each irreducible representation.
In Section~\ref{sec:likely}, we return to consideration of the likelihood function for elapsed time under a stochastic model and discuss possible future directions to efficiently evaluate this function using approximate, numeric means based solely on eigenvalue algorithms.
We close with a discussion of our plans for future explorations of the key ideas presented in this paper.

\section{Rearrangement models for bacterial genomes}
\label{sec:rearrangementmodels}
There is a diverse range of evolutionary events that cause rearrangements of regions on the bacterial genome.  
The rearrangements that are responsible for the majority of large-scale changes include \emph{inversion}, \emph{translocation}, \emph{duplication}, \emph{insertion} (through lateral gene transfer), and \emph{deletion} (for example, see~\cite{klippel1993analysis} or the reviews~\cite{grindley2006mechanisms,yang2010topoisomerases}). 
The techniques used in this paper are suited to those that yield rearrangement models based on permutation groups \cite{egri2014group,francis2014algebraic}, in which each event is invertible (able to be undone). 
These include inversions (the most frequent rearrangement event), as well as translocation.  
To simplify the discussion, we will focus on inversions.
An inversion refers to the excision of a segment of DNA, followed by its reinsertion in the same place but with the opposite orientation.  
These events are commonly performed by the action of ``site-specific recombinases".  
These enzymes act by cutting two strands that cross (at specific short sequences that the enzyme recognizes), and rejoining them to each other.

A common modelling simplification for these processes is to envisage the circular bacterial genome as consisting of a cyclic ordering of $N$ genes (or, more generally, any identifiable set of homologous\footnote{Evolutionary related.} regions of DNA).
We depict the canonical cyclic ordering in Figure~1.
In this picture, each individual gene/region can be considered as either (i) orientated, or (ii) unorientated (with directionality defined by the chemical orientation of a single strand of nucleic acid).
For the unoriented case the relevant group is the symmetric group $\mathcal{S}_N$, while in the oriented case the group is the hyperoctahedral group (these are the Coxeter groups of types $A$ and $B$ respectively, in the context of Lie theory).
In this paper we will focus on the the simpler unorientated formulation
because we are primarily interested in addressing the core computational obstructions inherent in all rearrangement models, and there is no advantage in further obscuring the discussion with more complicated (albeit, more realistic) models.
In any case, the algebraic ideas we present are easily adaptable to any group-based rearrangement model one cares to use.
See~\cite{egri2014group,francis2014algebraic} for descriptions of a range of possible models amenable to algebraic approaches.

\begin{figure}
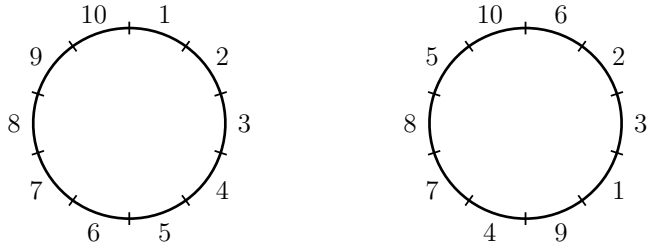

\label{fig:cyclicorder1}
\begin{center}

\resizebox{.25\textwidth}{!}{
\def\sp{{10,1,2,3,4,5,6,7,8,9,10}}\unsignedpermutation
}
\qquad
\resizebox{.25\textwidth}{!}{
\def\sp{{10,6,2,3,1,9,4,7,8,5,10}}\unsignedpermutation
}
\caption{Two circular genomes, each represented by an arrangement of $N\!=\!10$ regions. The ``canonical'' cyclic ordering is on the left.  }
\end{center}
\end{figure}

Given a genome as an (unorientated) cyclic order of $N$ genes, 
the effect of a rearrangement is described as the application of a permutation $\alpha\in\mathcal{S}_N$ understood as moving the gene in location $i$ to location $\alpha(i)$ (this is the ``positions'' paradigm as described in~\cite{bhatia2016position}).
Taking the canonical cyclic ordering as reference, each genome is determined by a permutation $\sigma\in \mathcal{S}_N$ where $j\!=\!\sigma(i)$ is understood as indicating that gene $i$ is in location $j\!=\!\sigma(i)$. 
We can then realise a sequence of rearrangements in mathematical terms as a composition (in the symmetric group\footnote{We compose permutations from right to left, so $(1,2)\circ (2,3)\!\equiv\! (1,2)(2,3)\!=\!(1,2,3)$, for instance.}) of a sequence of permutations.
That is, the rearrangement $\alpha$ converts the genome determined by $\sigma$ to the genome determined by $\alpha\circ \sigma\!\equiv\! \alpha \sigma$.

Without a fixed frame of reference for gene locations, we must treat as equivalent any two genomes that can be identified after rotation and/or reflection.
Given we only observe the relative location of each gene in a genome, it is appropriate to describe these equivalences using dihedral permutations $\mathcal{D}_N$, that is, the group of symmetries generated by reflection and cyclic rotation on $N$ elements.
To implement this, we declare two permutations $\sigma, \sigma'\in\mathcal{S}_N$ to determine equivalent genomes if there is a permutation $d\!\in\! \mathcal{D}_N$ such that $\sigma'\!=\!d\sigma$.
This is reasonable, since $k\!=\!d(j)\!=\!d\sigma(i)\!=\!\sigma'(i)$ is understood as indicting gene $i$ is in location $k\!=\!d(j)$.
When a frame of reference is not fixed, this means we may identify each genome with a right coset $\mathcal{D}_N\sigma$, which is consistent with the orbit-stabilizer theorem in identifying the number of distinct genomes as $|\mathcal{S}_N/\mathcal{D}_N|=N!/2N$.

Defining a biological model of rearrangements $\mathcal{M}$ amounts to fixing a subset $\mathcal{M}\subset\mathcal{S}_N$, most commonly consisting entirely of involutions (self inverse permutations\footnote{All inversions are described by involutions, however the converse is false. For example, the permutation $(1,2)(3,4)$ is an involution but does not describe an inversion.}).
For example, perhaps the most simple yet biologically plausible model emerges by restricting to inversion of adjacent regions only, so
\beqn
\label{eq:adjacent}
\mathcal{M}=\{(1,2),(2,3),(3,4),\ldots,(N-1,N),(N,1)\}.
\eqn
On this point however we keep the discussion general and suppose the biological model consists of $R$ involutions: 
\[
\mathcal{M}=\{s_1,s_2,\ldots,s_R\}\subset \mathcal{S}_N
\]
so $s_i^2=e$ for each $i=1,2,\ldots, R$.

Given a genome $G_0$ and a fixed frame of reference, without loss of generality we may label the regions of $G_0$ canonically such that  $G_0\!\equiv\! e$, and the biological model is implemented by composing members of $\mathcal{M}$ with $e$.
For instance the sequence of rearrangements $s_{i_1},s_{i_2},\ldots,s_{i_k}$ produces the new genome $G\!\equiv\! \sigma$ where
\[
\sigma=s_{i_k}\ldots s_{i_{2}}s_{i_1}e=s_{i_k}\ldots s_{i_{2}}s_{i_1}.
\]

On top of the chosen biological model $\mathcal{M}$, we choose a simple stochastic model of genome evolution where rearrangements occur randomly in time with:
\begin{enumerate}[(i)]
\item Events governed by a Poisson distribution with fixed rate $r$;
\item Each event is a rearrangement chosen uniformly at random from $\mathcal{M}$. 
\end{enumerate}
We note that when $\mathcal{M}$ is closed under inverses (which is the case if $\mathcal{M}$ consists entirely of involutions), we have a reversible stochastic process.

As is quite normal in phylogenetic modelling, we cannot infer ``molecular rate'' $r$ independently from time elapsed $T$ without further calibration  \cite{penny2005} (from, for example, the fossil record).
Thus, for simplicity we set $r\!=\!1$ throughout and interpret $T$ in units of mean number of events.

Given this setting, the primary computational obstruction we explore --- and ultimately bypass --- is the calculation of the number of ways of rearranging one genome into another using $k$ rearrangements taken from the fixed, but arbitrary, biological model $\mathcal{M}$.
For each $\sigma\in\mathcal{S}_N$, this amounts to the calculation of the number $\beta_k(\sigma)$ of length $k$ words with letters $s_{i_j}$ taken from the alphabet $\mathcal{M}$ such that 
\[
\sigma = s_{i_k}s_{i_{k-1}}\ldots s_{i_1}.
\]

From the biological point of view, when considering two present-day genomes, we rely on homology to identify common regions and consider one as the canonically ordered reference $G_0\!\equiv\! e$ and the other as the target $G\!\equiv\! \sigma$. 
Suppose, further, we would like to obtain the evolutionary distance between these two genomes.
As discussed above, since we are assuming the biological model $\mathcal{M}$ consists entirely of involutions, our stochastic model is reversible and we may, without loss of generality, consider $G$ to be the descendant of $G_0$.  
However, under this assumption, we cannot be certain how to orientate $G$ with respect to $G_0$ under the dihedral symmetries $\mathcal{D}_N$ mentioned above.
These observations imply we should take
\beqn
\label{eq:beta}
\alpha_k(G_0\rightarrow G)\equiv\alpha_k(\sigma):=\sum_{d\in \mathcal{D}_N}\beta_k(d\sigma),
\eqn
as the number of passages from genome $G_0$ to the descendant $G$ (or equivalents) under $k$ rearrangements $s_i\in\mathcal{M}$. 

The likelihood for the passage of evolutionary time $T$ is defined as the conditional probability
\beqn
\label{eq:likely}
 L(T|G_0\rightarrow G)=\text{prob}(G_0\rightarrow G|T)=\sum_{k\geq 0}\frac{\alpha_k(G_0\rightarrow G)}{R^k}\frac{T^ke^{-T}}{k!},
\eqn
where, consistent with our assumption that each rearrangement is equally likely, $R^k$ is recognized as the total number of passages of length $k$.
One would then like to find the optimum of this function and hence obtain the maximum likelihood estimate $\widehat{T}$ of elapsed time. 
As we will see, the key computational obstruction to this ideal is the efficient calculation of the counts $\beta_k(\sigma)$.
In \cite{serdoz2016maximum} we found that the iterative algorithm described in \cite{clark2016bacterial} can be used to effectively compute (in a matter of hours) counts for genomes with a maximum of $N\!=\!9$ regions.



\section{Computation of path counts using group characters}
\label{sec:regrep}
In this section we show how to compute the counts $\beta_k(\sigma)$ using the irreducible representations of the symmetric group, ultimately showing the computation reduces to an eigenvalue problem. 

Consider the symmetric group $\mathcal{S}_N$ and a rearrangement model $\mathcal{M}=\{s_1,s_2,\ldots, s_R\}\subset \mathcal{S}_N$. 
Given a target genome $G\!\equiv\! \sigma$, the basic data we need to compute is $\beta_k(\sigma)$, the number of words of length $k$ equal to $\sigma$ using the letters $s_i\in \mathcal{M}$.
We define $s:=\sum_{i=1}^R s_i$, where the sum is taken in the group algebra $\mathbb{C}\mathcal{S}_N$ (formal linear combinations of group elements over the complex field) and observe that
\beqn
\label{eq:sGroupAlgebra}
s^{k}
=
(s_1+s_2+\ldots +s_R)^k
=\sum (\text{words of length $k$ in the $s_i$})
:=\sum_{\sigma\in \mathcal{S}_N}\beta_k(\sigma)\sigma.
\eqn

We take $\chi:\mathcal{S}_N\rightarrow \mathbb{C}$ as the group character of the associated regular representation $\theta:\mathcal{S}_N\rightarrow GL(N!,\mathbb{C})$ (obtained by letting $\mathcal{S}_N$ act on itself on the left).
Since there are no fixed points for this action,  $\chi(\sigma)\!=\!N!$ if $\sigma\!=\!e$ and is zero otherwise.
Thus we have (extending $\chi$ linearly to all of the group algebra):
\beqn
\label{eq:betak}
\beta_k(\sigma)=\frac{1}{N!}\text{tr}(\theta(\sigma^{-1}s^k))=\frac{1}{N!}\chi(\sigma^{-1}s^k).
\eqn 

Immediately it follows that, for each $f\in \mathcal{S}_N$ satisfying $f^{-1} sf\!=\!s$, we have equality of path counts $\beta_k(\sigma)\!=\!\beta_k(f^{-1}\sigma f)$ since (using the cyclic property of the trace): 
\[
 \chi((f^{-1}\sigma f)^{-1}s^k)=\chi\left(f\sigma^{-1}f^{-1}s^k\right)=\chi\left(\sigma^{-1}f^{-1}s^kf\right)=\chi\left(\sigma^{-1}(f^{-1}sf)^k\right)=\chi\left(\sigma^{-1}s^k\right).
\]
Consistent with \cite[Prop 5.1]{serdoz2016maximum}, in the specific case of the biological model (\ref{eq:adjacent})  this equality of path counts holds for all $f\in \mathcal{D}_N$, and, in general, the equality holds for all $f$ in the stabilizer of $s$ under the conjugation action $s\mapsto f^{-1}sf$.

Now suppose $\theta=\oplus_{\mu}d_\mu\theta_\mu$ is the decomposition of $\theta$ into irreducible representations $\theta_\mu:\mathcal{S}_N\rightarrow \text{GL}(d_\mu,\mathbb{C})$ indexed by integer partitions $\mu$ of $N$, with each irreducible appearing with multiplicity equal to its dimension $d_\mu$ \cite{sagan2001}.
This reduces the calculation of $\beta_k(\sigma)$ to computing, for each integer partition $\mu$, the trace of the matrix $\theta_\mu(\sigma^{-1}s^k)=\theta_\mu(\sigma^{-1})\theta_\mu(s)^k$.
That is,
\beqn
\label{eq:betaIrr}
\beta_k(\sigma)=\frac{1}{N!}\sum_\mu d_\mu\chi_\mu(\sigma^{-1}s^k)=\frac{1}{N!}\sum_\mu d_\mu\text{tr}\left(\theta_\mu(\sigma)^{-1}\theta_\mu(s)^k\right),
\eqn
where $\chi_\mu$ is the irreducible character corresponding to the irreducible representation labelled by $\mu$. 

Although this reduces the computation of $\beta_k(\sigma)$ to the calculation of some matrix products in each irreducible representation of the symmetric group, this procedure remains unfeasible for $N$ of reasonable size, since the dimension of the irreducible representations can still be very large. 
(For example, when $N\!=\!20$ --- which is hardly a biologically impressive case --- the maximum dimension over all irreducible representations is $\sim 10^9$.)

We further our analysis by noticing that if the biological model $\mathcal{M}$ consists entirely of involutions, that is $s_i^2=e$, then, in the regular representation, each $\theta(s_i)$ is a symmetric permutation matrix.
This in turn implies $\theta(s)=\sum_i^N\theta(s_i)$ is symmetric and hence diagonalizable (with real eigenvalues) and hence each irreducible representation $\theta_i(s)$ is also diagonalizable.

Recalling that a linear operator is diagonalizable if and only if its minimal polynomial $q$ is a product of distinct linear factors, the minimal polynomial of $s$ must have the form 
\[
q=(x-\lambda_1)(x-\lambda_2)\ldots (x-\lambda_D)
\]
for distinct eigenvalues $\lambda_i\!\in\! \mathbb{R}$.
The degree $D$ of the minimal polynomial of $\theta(s)$ will of course depend on the particular model chosen but we can at least say (via the Cayley-Hamilton theorem) that it is less than $N!$, since $|\mathcal{S}_N|=N!$ is the dimension of the regular representation.


For each irreducible $\theta_\mu$ it is also the case that $q(\theta_\mu(s))\!=\!0$, which in turn implies that the minimal polynomial $q_\mu$ of each irreducible $\theta_\mu(s)$ must divide $q$.
We again conclude (via the Cayley-Hamilton theorem) that $D_\mu:=\text{deg}(q_\mu)\leq d_\mu$.

For each $\mu$, we label the distinct eigenvalues of $\theta_\mu(s)$ as $\lambda^{(\mu)}_1,\lambda^{(\mu)}_2,\ldots,\lambda^{(\mu)}_{D_\mu}\in\mathbb{R}$ and  define, for each $a\!=\!1,2\ldots,D_\mu$, the projection operators $E_{a}^{(\mu)}$ onto the eigenspaces corresponding to eigenvalue $\lambda^{(\mu)}_a$ via
\[
E_{{a}}^{(\mu)}=\prod_{b \neq a}\frac{\theta_\mu(s)-\lambda_{b}^{(\mu)}}{\lambda^{(\mu)}_{a}-\lambda^{(\mu)}_{b}}.
\] 
By construction, for each irreducible $\mu$, the $E^{(\mu)}_{a}$ provide a set of orthogonal projector operators:
\[
\sum_{a=1}^{D_\mu}E^{(\mu)}_{a}=\id,\qquad E^{(\mu)}_{a}E^{(\mu)}_{b}=\delta_{ab}\id,
\]
where $\delta_{ab}$ is the Kronecker delta (equal to 1 if $i\!=\!j$ and 0 otherwise).

These observations allow us to write, for each $\mu$:
\beqn
\label{eq:projform}
\theta_{\mu}(s)^k=\sum_{a=1}^{D_\mu}(\lambda_{a}^{(\mu)})^kE^{(\mu)}_{a}
\eqn
and hence
\[
\text{tr}\left(\theta_\mu(\sigma)^{-1}\theta_\mu(s)^k\right)=\sum_{a=1}^{D_\mu}(\lambda_{a}^{(\mu)})^k\text{tr}\left(\theta_\mu(\sigma)^{-1}E_a^{(\mu)}\right)=\sum_{a=1}^{D_\mu}(\lambda_{a}^{(\mu)})^k\text{tr}\left(E_a^{(\mu)}\theta_\mu(\sigma)^{-1}E_a^{(\mu)}\right),
\] 
where in the last equality we use the cyclic property of the trace and orthogonality of the projection operators $E_a^{(\mu)}$. 

Recall that each irreducible representation $\theta_\mu$ can be associated with the action of $\mathcal{S}_N$ on an irreducible $\mathcal{S}_N$-module $V^\mu\cong \mathbb{C}^{D_\mu}$ \cite{sagan2001}. 
Our results thus far reduce the computation of $\beta_k(\sigma)$ to
\begin{enumerate}[(i)]
\item Computation of the eigenvalues $\{\lambda^{(\mu)}_a\}$ of $s$ on each irreducible module $V^\mu$ under this action;
\item For each eigenvalue $\lambda^{(\mu)}_a$ and eigenspace $\text{Eig}(s;\mu,a)\!=\!E_a^{(\mu)}V^\mu$, computation of the eigenvalues of $\theta_\mu(\sigma)^{-1}$ restricted to $\text{Eig}(s;\mu,a)$ under $\theta_\mu(\sigma)^{-1}\mapsto E^{(\mu)}_a\theta_\mu(\sigma)^{-1}E^{(\mu)}_a$.
\end{enumerate}

As a simple first example, we use these ideas to compute exact, closed forms of $\alpha_k(\sigma)$ for $N\!=\!4$ under the biological model $\mathcal{M}=\{(1,2),(2,3),(3,4),(1,4)\}$.
In this case, we have $s=(1,2)+(2,3)+(3,4)+(1,4)\in \mathcal{S}_4$ and the regular representation $\theta(s)$ can be computed by hand without too much trouble as an explicit $24\times 24$ matrix.
We find that $\theta(s)$ has minimal polynomial
\[
q=x(x-2)(x+2)(x-4)(x+4).
\]
The irreducible representations of $\mathcal{S}_4$ are labelled by the integer partitions $\{4,31,2^2,21^2,1^4\}$, and the character table is given in Table~1.

\begin{table}
\label{tab:s4characters}
\begin{centering}
\begin{tabular}{c|ccccc}
 & [e] & [(1,2)] & [(1,2)(3,4)] &  [(1,2,3)] & [(1,2,3,4)] \\
 \hline
 (4) & 1 & 1 & 1 & 1 & 1 \\
 (31) & 3 & 1 & -1 & 0  & -1 \\
 ($2^2$) & 2 & 0 & 2 & -1 & 0 \\
 ($21^2$) & 3 & -1 & -1 & 0  & 1 \\
 ($1^4$) & 1 & -1 & 1 & 1 & -1 
\end{tabular}
\caption{The character table of $\mathcal{S}_4$. The rows are labeled by the irreducible representations, the columns by conjugacy classes. Note that $(4)$ and $(1^4)$ label the trivial and sign representations, respectively.}
\end{centering}
\end{table}


The corresponding minimal polynomials are given by
\beqn
q_{4}=(x-4),\quad q_{31}=x(x-2),\quad q_{2^2}=(x-2)(x+2), \quad q_{21^2}=x(x+2),\quad q_{1^4}=(x+4).\nonumber
\eqn
(The answers for the trivial and sign representations are immediate given $\sgn{s_i}=-1$ for each $s_i\in \mathcal{M}$.)
Using the general formula (\ref{eq:projform}) we have, for $k>0$:
\beqn
\theta_{4}(s^k)&=4^k;\qquad \theta_{1^4}(s^k)=(-4)^k;\quad \nonumber
\theta_{2^2}(s^k)=2^k\fra{1}{4}\theta_{2^2}(s+2)+(-2)^k(-\fra{1}{4})\theta_{2^2}(s-2);\\
\theta_{31}(s^k)&=0^k(-\fra{1}{2})\theta_{31}(s-2)+2^k\fra{1}{2}\theta_{31}(s);\quad 
\theta_{21^2}(s^k)=0^k(\fra{1}{2})\theta_{21^2}(s+2)+(-2)^k(-\fra{1}{2})\theta_{21^2}(s).
\eqn
After simplifying and with reference to the character table, for $k>0$ odd we obtain  the (constant time in $k$) expressions
\beqn
\chi(\sigma^{-1}s^k)&=4^k(1-\text{sgn}(\sigma))+3\cdot 2^{k-1}\chi_{31}(\sigma^{-1}s)+2\cdot 2^{k-1}\chi_{2^2}(\sigma^{-1}s)+3\cdot 2^{k-1}\chi_{21^2}(\sigma^{-1}s),\nonumber
\eqn
where we have used the decomposition of the regular representation into irreducibles  
\[
\chi=\chi_{4}+3\chi_{31}+2\chi_{2^2}+3\chi_{21^2}+\chi_{1^4}.
\]
For $k>0$ even, we similarly obtain
\beqn
\chi(\sigma^{-1}s^k)&=4^k(1+\text{sgn}(\sigma))+3\cdot 2^{k-1}\chi_{31}(\sigma^{-1}s)+2\cdot 2^{k}\chi_{2^2}(\sigma^{-1})-3\cdot 2^{k-1}\chi_{21^2}(\sigma^{-1}s).\nonumber
\eqn

Dividing each of these by $4!$ and summing over the dihedral group as in (\ref{eq:beta}), we find explicit expressions (constant time in $k$) for the number of passages $G_0\rightarrow G\equiv \sigma$ using rearrangement moves belonging to $\mathcal{M}=\{(1,2),(2,3),(3,4),(4,1)\}$.
In particular we see that we have the limits (as a proportion of total number of passage counts):
\[
\frac{\alpha_k(\sigma)}{4^k}\rightarrow 
\left\{ 
\begin{array}{l}
\fra{1}{12} \text{ if $\sgn{\sigma}=\pm 1$ and $k$ is even/odd, respectively,}\\
0,\text{ otherwise;}
\end{array}
\right.
\]
simply reflecting that this particular model $\mathcal{M}$ consists entirely of odd involutions.




\section{Calculation of likelihood function}
\label{sec:likely}
Here we use the results of the previous section to further our goal of finding an efficient (possibly approximate) method for evaluating the likelihood function (\ref{eq:likely}). 
Firstly, we continue our description using the group algebra (and regular representation) to identity our model as a continuous-time Markov chain with associated rate matrices arising from a `group-based' model (as coined in the phylogenetics literature \cite{semple2003}).
This discussion furthers our understanding of the model and guides the way to significant generalization in which different rearrangements may occur at different (Poisson distributed) rates.
Secondly we discuss the decomposition of the likelihood function  (as a function on the symmetric group) into contributions from each irreducible part.

As discussed in Section~\ref{sec:rearrangementmodels}, we suppose mutation events are Poisson distributed in time with rate $r$ and are chosen uniformly at random from a model $\mathcal{M}$, consisting of $R\!=\!|\mathcal{M}|$ allowed rearrangements (as above, we again set $r\!=\!1$ and interpret $T$ in units of mean number of events).
Interpreting genome $G\!\equiv\! \sigma$ as the descendant $G_0\!\equiv\! e$, using (\ref{eq:beta}) and (\ref{eq:betak})  we can express the likelihood function (\ref{eq:likely})  as
\beqn
L(T|\sigma)\equiv L(T|G_0\rightarrow G)=\sum_{k\geq 0}\frac{\alpha_k(\sigma)}{R^k}\frac{T^ke^{-T}}{k!}
&=\frac{1}{N!}\sum_{k\geq 0}\sum_{d\in \mathcal{D}_N}\frac{\chi\left(\sigma^{-1}d^{-1}s^k\right)}{R^k}\frac{T^ke^{-T}}{k!}\\
&=\frac{1}{N!}\sum_{k\geq 0}\sum_{d\in \mathcal{D}_N}\frac{\chi\left(\sigma^{-1}ds^k\right)}{R^k}\frac{T^ke^{-T}}{k!}.\nonumber
\eqn
Rescaling time as $T \rightarrow T/R$, we find
\beqn
L(T|\sigma)&=e^{-RT}\frac{1}{N!}\sum_{k\geq 0}\sum_{d\in \mathcal{D}_N}\chi(\sigma^{-1}ds^k)\frac{T^k}{k!}\nonumber\\
&=e^{-RT}\frac{1}{N!}\sum_{d\in \mathcal{D}_N}\chi\left(\sigma^{-1}d\sum_{k\geq 0}\frac{T^k}{k!}s^k\right)\\
&=e^{-RT}\frac{1}{N!}\sum_{d\in \mathcal{D}_N}\chi\left(\sigma^{-1}de^{sT}\right)\\
&=\frac{1}{N!}\sum_{d\in \mathcal{D}_N}\chi\left(\sigma^{-1}de^{(s-R)T}\right),
\eqn
where $R\equiv Re\in \mathbb{C}\mathcal{S}_N$ is understood as a scaling of the identity element in the group algebra.

Since each $s_i\in\mathcal{M}$ occurs as a permutation matrix in the regular representation, we observe that $Q:=\theta(s-Re)=\theta(s_1+s_2+\ldots+s_N-Re)$ is a ``rate matrix'' (zero-column sums) associated to a continuous-time Markov chain.
Thus our likelihood function can be written in an elegant form as the character of an analytic function on group algebra:
\[
L(T|G_0\rightarrow G)=\frac{1}{N!}\sum_{d\in \mathcal{D}_N}\chi(d\sigma^{-1}e^{QT}).
\]

From this point of view $e^{QT}$ is the homogeneous, time-dependent probability transition matrix whose $ij$ entry gives the probability of a transition from the $j^{\text{th}}$ group element to the $i^{\text{th}}$ group element in time $T$.
This shows that our model is a special case of a so-called ``group-based'' model \cite{semple2003} (usually formulated for abelian groups only\footnote{See \cite{sumner2011} for construction of a group-based model using the regular representation of any finite group.}), where a more general, rate matrix would naturally be given as
\[
Q=-\widehat{\gamma}\id+\sum_{f\in \mathcal{S}_N}\gamma_{f}\theta(f),
\]
with $\widehat{\gamma}=\sum_{f\in \mathcal{S}_N}\gamma_f$ for some parameters $\gamma_f\geq 0$ giving the rate of transition of group elements $\sigma\rightarrow f^{-1}\sigma$.

From this point of view, one can think of our stochastic model as following from the biological model $\mathcal{M}$ together with the special choice $\gamma_f\!=\!1$ when $f\in \mathcal{M}$ and $\gamma_f\!=\!0$ otherwise.
Treating each $\gamma_f$ as an independent modelling parameter is a nice generalization, but we do not explore it further beyond making the point it seems unlikely one would be able infer values of the rates $\gamma_f$ from observed data given only two genomes and their relative gene arrangement as input data.

Returning to the likelihood function, we see that 
\[
L(T|\sigma)=\sum_{\mu}d_\mu L_\mu(T|\sigma),
\]
where each $L_\mu(T|\sigma)$ is a contribution to the likelihood function corresponding to each irreducible representation of the symmetric group:
\[
L_\mu(T|\sigma):=\frac{1}{N!}\sum_{d\in \mathcal{D}_N}\chi_\mu(\sigma^{-1}de^{QT}).
\]
In turn, each summand in this expression can be computed, using the orthogonal projection operators $E_a^{(\mu)}$, as
\[
\chi_\mu(\sigma^{-1}de^{QT})=e^{-RT}\sum_{a=1}^{D_\mu}e^{\lambda_a^{(\mu)}T}\text{tr}(\theta(\sigma^{-1}d)E_a^\mu).
\]

This illustrates that, in general terms, each eigenvalue $\lambda$ of the operator $s$ contributes to the likelihood function via the substitution $\lambda \rightarrow e^{\lambda T}$.
Hence, by far the most significant contributions to the likelihood function are going to come from the largest eigenvalues.
We hope that this observation will lead to development of an efficient, approximate method for computing the maximum likelihood estimate of time elapsed under general genome rearrangement models. 

\section{Discussion}
In this paper we have explored the application of the representation theory of finite groups to the problem of computing evolutionary distance between bacterial genomes.
In Section~\ref{sec:rearrangementmodels} we showed how to construct genome rearrangement models in the context of permutation groups and gave the likelihood function for time elapsed under a simple stochastic model based on Poisson distributed events (\ref{eq:likely}). 
In Section~\ref{sec:regrep} we reformulated the problem of counting the number of passages from one genome to another (under a selected set of possible rearrangements) into the calculation of coefficients in the group algebra of the symmetric group (\ref{eq:sGroupAlgebra}).
We then interpreted these coefficients through a trace calculation on each irreducible representation of the symmetric group (\ref{eq:betaIrr}).
Finally, in Section~\ref{sec:likely} we showed that, if one is primarily interested in maximum likelihood estimates of elapsed time, it is possible to sidestep computation of these counts and evaluate the likelihood function using the eigenvalues of the matrices representing the group algebra element $s$ under each irreducible representation.

Our future directions will focus on this final observation, particular in the context of numerical approximations.
For instance, we intend to couple combinatorial constructions of the irreducible representations of the symmetric group (such as the Specht modules \cite{sagan2001}) with standard eigenvalue algorithms (such as Lanczos iteration \cite{trefethen1997}).
The outcomes of these explorations will be the subject of future work.

\bibliographystyle{plain}
\bibliography{masterC}

\end{document}